\magnification=\magstep1
\baselineskip=14pt
\font\titulobold=cmbx12 scaled\magstep1
%\font\twelverm=cmr12
 
\  {\hfill \bf GDNL\#3/97}

{\hfill \bf Revised version}
\bigskip
\bigskip 

\titulobold
\bigskip
\centerline{Validation and Calibration of Models}
\centerline{for Reaction-Diffusion Systems}

\tenrm
\bigskip

\centerline{\bf Rui Dil\~ ao\footnote{\rm $^1$}{\rm rui@sd.ist.utl.pt} {\rm and} \bf Joaquim Sainhas\footnote{\rm $^2$}{\rm jsainhas@sd.ist.utl.pt}}
\bigskip

\centerline{\it Grupo de Din\^amica N\~ao-Linear, Departamento de F\'\i sica\footnote{}{\rm Phone: (351)-(1)-8417617, FAX: (351)-(1)-8419123}}
\centerline{\it Instituto Superior T\'ecnico, Av. Rovisco Pais}
\centerline{\it 1096 Lisboa Codex, Portugal}

\bigskip
%\bigskip
%\bigskip
\centerline{\bf Abstract}
Space and time scales are not independent in diffusion. In fact, 
numerical simulations show that different patterns are obtained when
space and time steps ($\Delta x$ and $\Delta t$) are varied independently. 
On the other hand, anisotropy effects due to the symmetries of the 
discretization lattice prevent the quantitative calibration of models.
We  introduce a new class of explicit difference methods 
for numerical integration of diffusion and reaction-diffusion  
equations, 
where the dependence on space and time scales occurs naturally.
Numerical solutions approach the exact solution of the continuous diffusion
equation  for finite $\Delta x$ and $\Delta t$, if the  parameter $\gamma_N=D \Delta t/(\Delta x)^2$ assumes a fixed constant 
value, where $N$ is an odd positive integer parametrizing the alghorithm. 
The error between  
the solutions of the discrete and the continuous equations goes to zero as 
$(\Delta x)^{2(N+2)}$ and the values of $\gamma_N$  are 
dimension independent. With these new integration methods,   
anisotropy effects resulting from the finite differences 
are minimized,  defining a standard 
for validation and calibration  of numerical solutions of diffusion and
reaction-diffusion equations. 
Comparison between numerical and analytical solutions of reaction-diffusion
equations give global discretization errors of the order of $10^{-6}$ in the {\it sup} norm. 
Circular patterns of travelling waves have a maximum 
relative random deviation from the spherical symmetry of the order of $0.2\%$,
and the standard deviation of the fluctuations around the mean circular wave
front is of the order of $10^{-3}$.

\bigskip
\bigskip

\vfill\eject

{\bf 1 - Introduction}
\medskip

After the important work of Turing [1952] on the chemical basis of 
morphogenesis, theoretical and experimental studies have shown that solutions of reaction-diffusion (RD) partial differential equations present
 self-organizing properties,
common to biological and chemical extended systems. These  properties
appear as emerging coherent patterns or structures in spatially 
extended media.
Due to the nonlinear nature of local kinetic me\-cha\-nisms in RD systems, local fluctuations can grow and propagate by diffusion to the surrounding media. 
For this reason, extended media
supporting local nonlinear kinetic mechanisms are called excitable media
[Winfree, 1990; Keener \& Tyson, 1992]. 

The prototype experimental system for the study of pattern formation
in  extended media are the Belousov-Zhabotinsky reaction [Zaikin \& Zhabotinsky, 1970; Zhabotinsky \& Zaikin, 1973] and,
in biological systems, the aggregation patterns of colonies of
{\it Dictyostelium discoideum}, [Tyson {\it et al.}, 1989; Steinbock 
{\it et al.}, 1993]. 
Due to the similarity of dynamic behaviors, these experimental systems
provide insights   into the mechanisms of pattern formation (morphogenesis).
%one the most challenging research issues in the study of patterns 
%in biological systems. 
A large collection of experimental
evidence of spatiotemporal patterns  in nature can be found in 
Field, K\"or\"os \& Noyes [1972],
Meinhardt  [1982],  Harrison [1993], Cross \& Hohenberg [1993], 
Kock  \& Meinhardt  [1994] and Epstein \& Showalter [1996].

Reaction-diffusion (partial differential) equations are the simplest mathematical 
models for the study of pattern formation in excitable media.
The coherent patterns appearing in numerical simulations of RD equations are
undamped travelling  waves, spiral travelling  waves, 
stable strips and spotty structures (Turing patterns), 
[Castets {\it et al.},1990].  From the qualitative point of view, 
the same structures
can be obtained with cellular automata models, [Markus \& Hess, 1990;
Durrett \& Griffeath, 1993]. However, 
cellular automata models cannot be calibrated and validated from real parameters (kinetic reaction rates and diffusion coefficients),
as the physical and chemical mechanisms that characterize the observed phenomena are generally continuous models. 
These parameters are specially important as different pattern topologies
are obtained when they are varied (bifurcation phenomena).

The possibility to validate a physical or chemical mechanism
for the emergence of a given coherent pattern lies on the existence of
reliable numerical alghorithms.
 If simulation models show
the same qualitative behavior  as real systems, model parameters can be
calibrated   with experimental data. If the calibrated model has
the same behavior as the real system, the model is validated, and
the association of specific parameter
values to the real system characterizes its properties, 
[Oreskes {\it et al.}, 1994]. 
A classical example of this methodology is obtained in the study of
diffusion. In this case, the diffusive properties of a medium 
are quantified by the diffusion coefficient, measured by comparing experimental concentration
profiles  with the analytical solutions of the diffusion equation. Therefore,
diffusion, originated by the random impacts of the solvent molecules with suspended particles, 
is quantifiable by the diffusion coefficient, measuring the mean square
displacement of the suspended particles  [Chandrasekhar, 1943].

As experimental RD systems  are in general nonlinear, the 
calibration and validation of models should be based  on the
comparison between  the data obtained 
with  numerical  solutions of RD  equations and
data acquired in experiments.  This rises the problem of
 comparison between numerical and 
analytical solutions of nonlinear equations --- bench-marking.
 
\bigskip
{\bf 1.1 - Problems with numerically obtained  patterns with RD equations}
\medskip

The prototype mathematical model for the study of pattern formation in excitable media
is the two dimensional system of RD  equations
$$
{\partial \vec \varphi \over \partial t} = \vec X (\vec \varphi) + \vec D^T 
\cdot
\left( {\partial^2 \vec \varphi \over \partial x^2} + 
{\partial^2 \vec \varphi \over \partial y^2} \right)\eqno(1.1)
$$
where $\vec \varphi = (\varphi_1 ,\varphi_2)$ are the concentrations 
of some   chemical species or morphogens [Turing, 1952],
$\vec D^T = (D_1, D_2)$ are  diffusion coefficients and 
$\vec X^T=(X_1,X_2)$ is a vector
field describing the local kinetics of the system. Systems of partial
differential equations of this type describe, for example, the space-time 
pattern formation  in the Belousov-Zhabotinsky reaction and 
the interaction between cells separated by permeable membranes, [Turing, 1952;
Zhabotinsky \& Zaikin, 1973; Murray, 1993].  

In general, for non-linear vector fields, there are no general analytical 
techniques to obtain solutions of (1.1) and we must rely on numerical methods.
Stability and consistency conditions for finite difference 
integration methods  insure  that 
the truncation errors of   numerical solutions vanish when
the discrete space and time steps go simultaneously to zero [Smith, 1985; 
Sewell, 1988]. 
However, for large integration
times the accumulated error can grow indefinitely [John, 1971],
distorting  concentration patterns.

The simplest finite difference explicit method used in  numerical integration
of eq. (1.1) is the system of difference equations
$$\eqalign{
\varphi_{1,i,j}^{t+\Delta t} &=\Delta t X_1(\varphi_{1,i,j}^t,\varphi_{2,i,j}^t )+  
{D_1\Delta t \over (\Delta x)^2} 
\left( \varphi_{1,i-1,j}^t + \varphi_{1,i+1,j}^t + \varphi_{1,i,j-1}^t +
\varphi_{1,i,j+1}^t -4\ \varphi_{1,i,j}^t\right)\cr
\varphi_{2,i,j}^{t+\Delta t} &=\Delta t X_2(\varphi_{1,i,j}^t,\varphi_{2,i,j}^t )+ 
{D_2 \Delta t \over (\Delta x)^2} 
\left( \varphi_{2,i-1,j}^t + \varphi_{2,i+1,j}^t + \varphi_{2,i,j-1}^t +
\varphi_{2,i,j+1}^t -4\ \varphi_{2,i,j}^t\right)\cr} \eqno(1.2)
$$
where the subscripts $(i,j)$ refer to the coordinates  $(i \Delta x,j \Delta x)$ of the vertices of a  squared symmetric lattice in the plane.
The solutions at
time $t$, $\varphi_{1,i,j}^t$ and $\varphi_{2,i,j}^t$, are obtained iteratively through given initial 
functions $\varphi_1 (x,y,0)=f_1(x,y)$ and $\varphi_2 (x,y,0)=f_2(x,y)$,  
$\varphi_{1,i,j}^{0}=f_1(i \Delta x ,j \Delta x)$ and 
$\varphi_{2,i,j}^{0}=f_2(i \Delta x ,j \Delta x)$.
The centered difference scheme (1.2) is stable if 
$\gamma = \max \{D_1,D_2\}\Delta t/(\Delta x)^2 \le 1/4$, and the spatial truncation error 
is of the order of $(\Delta x)^2$,  [Smith, 1985; Sewell, 1988].

In the literature on numerical methods,
it is generally accepted that 
for the integration of two-dimensional diffusion and RD 
equations, the parameter $\gamma=D\Delta t/(\Delta x)^2$ 
must be as close as possible  to the value $\gamma =1/4$, 
[Carslaw \& Jaeger, 1959; Crank, 1975; Smith, 1985; Sewell, 1988]. 
This is justified by the simple fact that computation time is faster and
the choice of $\gamma $ is not very important if the purpose is to determine equilibrium solutions  [Press {\it et al.}, 1989].
In fact, if $\vec X=\vec 0$, equilibrium solutions of the difference scheme (1.2) coincide with the equilibrium solution of (1.1).  However, if the goal  is to obtain information about propagation 
velocities of diffusion fronts, or to study the pattern formation in 
reaction-diffusion processes, this approach presents serious problems, as 
we show now.

The first observation  is that numerically computed concentration profiles obtained with  RD  equations 
depend strongly on the magnitude of $\gamma =D\Delta t/(\Delta x)^2$, 
within the stability condition $\gamma \le 1/4$. To show this, we take
the  Brusselator model, [Prigogine \& Lefevre, 1968], 
in an extended two dimensional media, defined by the  equations
$$\eqalign{
{\partial  \varphi_1 \over \partial t} &=  
k_1 A-(k_2B+k_4)\varphi_1+k_3 \varphi_1^2\varphi_2+D_1
\left( {\partial^2   \varphi_1 \over \partial x^2} + 
{\partial^2   \varphi_1 \over \partial y^2} \right)\cr
{\partial  \varphi_2 \over \partial t} &=
k_2B\varphi_1-k_3\varphi_1^2\varphi_2 +D_2
\left( {\partial^2   \varphi_2 \over \partial x^2} + 
{\partial^2   \varphi_2 \over \partial y^2} \right)\cr }\eqno(1.3)
$$
where the kinetic terms correspond to the system of autocatalitic reactions
$A\to^{k_1} \varphi_1$, $B+\varphi_1\to^{k_2} \varphi_2+D$, $2 \varphi_1+\varphi_2\to^{k_3} 3\varphi_1$ and $\varphi_1\to^{k_4} E$,
the $k_i$'s are the reaction rates, and the concentrations $A$ and $B$ are 
kept constant. System (1.3) has an equilibrium unstable
solution for $\varphi_1=Ak_1/k_4$ and $\varphi_2=Bk_2k_4/(Ak_1k_3)$.

In Fig.  1,  different evolved patterns are shown, obtained with numerical integration method (1.2),
for the same initial condition on a lattice of $200\times 200$ cells and
several values of $\gamma= \max \{D_1,D_2\}\Delta t/(\Delta x)^2$.
The initial condition  has been chosen to produce spiral patterns,
[Klevecz {\it et al.}, 1992]. 
As $\gamma $ increases  within
the stability condition $\gamma \le 1/4$, patterns change shape and, after some
threshold, the spiral disappears leading to  concentric waves. 
For small values of $\gamma $,  
the actualization process induces a squared spurious symmetry in the concentration profiles, showing spatial anisotropy. 
These simulations show that
different nonequilibrium patterns  are obtained without 
changing either the physical parameters  (diffusion and kinetic coefficients),
or the initial conditions.  On the other hand, as 
$\Delta x=\sqrt{D\Delta t/ \gamma}$, different lattice lengths are obtained as
we vary $\gamma $, for the same number of lattice sites.

Therefore, the propagation velocity and physical dimensions of patterns 
obtained  numerically, as well as its symmetry properties,
are strongly dependent on the magnitude of $\gamma$,   leading to 
deformations in the patterns, together with quantitative changes
in the propagation velocities of wave fronts. 

On the other hand, the {\it ad hoc}  scaling introduced in Fig. 1, show that
space and time variables are not independent. This is a well known
symmetry associated with the diffusion equation, which is invariant under the
linear substitutions $x'=a x$ and $t'=a^2 t$ [John, 1971].

So, we are faced with the problem of knowing which of the simulations of
Fig. 1, better approximates  the solution of the  continuous
RD equation, for the prescribed initial data.

To avoid the problem of  numerical diffusion anisotropy in 
patterns of RD systems, some authors use 
implicit integration schemes as the Crank-Nicholson or ADI methods, [Press 
{\it et al.}, 1989],
which are computer time consuming. With explicit integration discrete methods,
Barkley [1990] proposed an 
explicit nine point formula difference approximation for the Laplace 
operator,  Pearson [1995] adds random noise, and Weimar {\it et al.} [1992] analyse several ``masks" for
the discrete approximation to the diffusion equation.
In the context of ecological modelling, Dejak {\it et al.} [1987] makes 
an error analysis as a function of the scaling parameter $\gamma$.
For cellular automata models, Markus \& Hesse [1990] use    
a random algorithm. All these strategies to overcome the problem of
diffusion anisotropy are heuristic and   estimates of global numerical 
errors are lacking. On the other hand, these   methods do not show  the role
of the heuristic scaling relation introduced in depicting the simulations 
of Fig. 1, which otherwise change the physical dimensions of the system.
For a general discussion on ``plausible-looking results which are qualitatively
incorrect"  see Ruuth [1995].

Another  problem  when simulating diffusion processes with explicit methods
is related to  the maximum propagation velocity of ``suspended particles".
For the discrete method $(1.2)$, in one time step, local concentration changes are due to next neighbors contributions and the maximum propagation velocity
is $v_{max} = 2{\Delta x\over \Delta t}$. But,
for the diffusion equation, $v_{max} = +\infty$, as  is well know. 
The solution of the discrete explicit method (1.2) converges to the solution
of Eq. (1.1), if  
$\lim_{\Delta t \to 0} \lim_{\Delta x \to 0} \gamma =$constant  
[John, 1971]. But as $\gamma =D\Delta t/(\Delta x)^2$, this last condition
implies that
$\lim_{\Delta t \to 0} \lim_{\Delta x \to 0} v_{max} = +\infty$, but for finite
$\Delta x$ and $\Delta t$, $v_{max}$ remains finite. Therefore,  
the best choice of $\gamma $ remained  an open question. 

According to  the microscopic interpretation of diffusion by the 
theory of Brownian motion, the  value of $\gamma$ should be, 
$\gamma = 1/2$, $\gamma = 1/4$ and $\gamma = 1/6$ in one, two and three
dimensions, respectively [Murray, 1993]. So, it seems natural to choose
one of the above values of $\gamma$ in simulations. However, these values
correspond to the upper limit of stability of explicit methods in one, two 
and three dimensions, respectively, and introduce numerical instabilities
due to roundoff effects.

\vfill \eject
{\bf 1.2 - An overview of results and organization of the paper}
\medskip

In this paper, we  derive a new class of explicit difference  methods 
for the numerical integration of diffusion and reaction-diffusion 
equations and a rigorous criterion for the choice of the 
space-time scaling parameter $\gamma = {D \Delta t/ (\Delta x)^2}$ in order that:
\item{a)} Explicit difference methods 
for the numerical integration of RD   equations
give accurate results for the transient solutions of diffusion and reaction-diffusion processes, enabling
to calibrate  model parameters with experimental data. 
\item{b)} The relation between space and time scales, 
$\lim_{\Delta t \to 0} \lim_{\Delta x \to 0} \gamma =$constant, 
a property of diffusion equation, is correctly interpreted.
\item{c)} The spatial distribution of concentrations do not depend on the
symmetries of the discretization lattice. In particular, diffusion anisotropy
in dimensions two and three are minimized, without additional hypotheses or
data modification. 

The calibration of the space and time scales 
results from a choice of a constant value for $\gamma$, dependent on
the  order of the global discretization error: $\gamma\equiv \gamma_N$, 
for an error of the order of $(\Delta x)^{2(N+2)}$, where $N\ge 1$ is an odd
integer.

The main consequences of the approach developed here are:
\item{i)} There exists an optimum value of $\gamma$, $\gamma=\gamma_N$, for
which the new class of explicit
difference methods shows the smallest global error,
when its solutions are compared with the solutions of the continuous 
diffusion and RD equation --- Fig. 2. The value
of $\gamma_N$ is dimension independent, defining
a space-time scaling relation.
\item{ii)} The algorithm is explicit and the accuracy of the simulations
can be arbitrarily increased, for finite $\Delta x$ and $\Delta t$ --- 
Eqs. (2.6), (2.34) and (2.47).
\item{iii)}  Numerical  lattice anisotropy in diffusion and reaction-diffusion
equations are minimized --- Fig. 5.
\item{iv)} In the context of the probabilistic interpretation of diffusion through Brownian
motion, the transition probabilities from one lattice cell to the adjacent
cells
is exactly calculable, being dependent on the number of 
neighborhood cells  taken into account in the discretization. 
The values of the transition probabilities determine the optimum Laplacian 
mask operator [Weimar {\it et al.}, 1992], as a function of neighborhood order. 
\item{v)}  Solutions of RD   equations with cylindrical 
or spherical symmetries numerically coincide, independently of the space
dimension of simulations --- Fig. 8.

\bigskip

In the next section we introduce the main technique for the discretization
of the diffusion equation in dimensions 1, 2 and 3, in order to account
for lattice anisotropies. The main results of the paper are stated in Theorems
A, B and C, and  the exact solutions of a linear RD   
equation are compared with the numerically calculated solution
--- bench-marking.
In Sec. 3 we summarize the main results of the paper from the applied point
of view.

\bigskip
\bigskip

{\bf 2 - Main Results}
\medskip

{\bf 2.1 - One space dimension}
\medskip

We begin with the diffusion equation in one space dimension
$$
 {\partial \phi\over \partial t} = D {\partial^{2} \phi\over \partial x^{2}}
\eqno(2.1)
$$
and we take a lattice $\Sigma $ in the $(x,t)$ slab with $x\in {\bf R}$
and $0\le t\le T$. With $\Delta x$ and $\Delta t$ as the increments in the variables $x$ and $t$, the lattice $\Sigma $ has coordinates 
$(n\Delta x,k\Delta t)$, with $n\in {\bf Z}$ and $k\in {\bf N}$. If $\phi(x,t)$
is a solution of the parabolic equation (2.1), satisfying the initial condition $\phi(x,0)=f(x)$, this solution, restricted to $\Sigma$, will be denoted by $\phi_n^k=\phi(n\Delta x,k\Delta t)$.

As it is well known, for the initial condition $f(x)$, the  solution of
the diffusion equation (2.1) is 
$$
\phi(x,t)=\int_{-\infty}^{+\infty}  f(y){1\over \sqrt{4\pi D t}}
e^{-{(x-y)^2\over 4 D t}} dy \eqno(2.2)
$$
If $f(x)$ is a discontinuous and integrable, the function $\phi(x,t)$
is of class $C^{\infty} ({\bf R})$, for every $t>0$. So, without loss of generality, we consider that $\phi(x,t)$ is of  class $C^{\infty} ({\bf R})$.
On the other hand, if $f(x)$ has compact support, $\phi(x,t)\not=0$ for every
$t>0$ and finite $x$. This implies that diffusive propagation  has
infinite velocity.

In order to discretize the diffusion equation (2.1) on the lattice 
$\Sigma $, we introduce the usual 
finite difference approximation to the space and time derivatives,
$$\eqalign{
{\partial \phi\over \partial t}&\simeq {\phi(x,t+\Delta t)-\phi(x,t)\over \Delta t} \cr
{\partial^{2} \phi\over \partial x^{2}}&\simeq
{\phi(x-i\Delta x,t)+\phi(x+i\Delta x,t)-2\phi(x,t)\over i^2 (\Delta x)^2}\cr }
\eqno(2.3)
$$ 
where $i$ is some positive integer. To account for large propagation velocities
in the discretization of the diffusion equation, we rewrite (2.1) in the form
$$
{\partial \phi\over \partial t} = D \sum_{i=1}^N \alpha_i
{\partial^{2} \phi\over \partial x^{2}}
\eqno(2.4)
$$
where the $\alpha_i$ are non negative constants such that
$$
\sum_{i=1}^N \alpha_i=1 \eqno(2.5)
$$
Introducing the finite difference approximation (2.3) into (2.4), we obtain 
the discrete equation 
$$
v_n^{k+1} =v_n^{k}+ \gamma_N \sum_{i=1}^{N} {\alpha_i\over i^2}
\left( v_{n-i}^k + v_{n+i}^k -2 v_n^k\right) \eqno(2.6)
$$
where
$$
\gamma_N=D {\Delta t\over (\Delta x)^2}\eqno(2.7)
$$
The integer $N$ is the space width of the finite difference equation (2.6) and
the constants $\alpha_i$ measure the  connectivity strength between
each cell and  its neighbors, up to width $N$. 
With $v_n^0=\phi_n^0=f(n\Delta x)$, all the $v_n^{k}$, for $k\ge 0$, are 
determined recursively.

For $N=1$,
the relation between the recursive 
solutions of (2.6) and the solutions of the diffusion equation for the same
initial function $f(x)=\phi(x,0)$ is easily obtained through the maximum
principle. As a matter of fact,  
if $\gamma_N$ is a sufficiently small  constant, then 
$$
|||v(x,t)|||\le ||f(x)||\eqno(2.8)
$$
where
$$
||f(x)||= sup_{x} |f(x)| \qquad {\rm and} \qquad |||v(x,t)||| = sup_{x,t} |v (x,t)|
$$
is the {\it sup} norm taken for $x\in {\bf R}$ and $0\le t\le T$. Under these
conditions, if $\Delta x, \Delta t \to 0$, $v_n^k$ converges uniformly to the
solution   of the  diffusion equation.
In the case $N=1$, the uniform convergence is achieved under the stability 
condition 
$\gamma_1=D \Delta t / (\Delta x)^2\le 1/2$, (Lemma II, \S 7.2 of 
[John, 1971]).

For the general case of $N\ge 1$, we apply   Schwartz inequality to (2.6):
$$
||v_n^{k+1}||\ \leq \ ||v_n^k||. |1-2\gamma_N \sum_{i=1}^N {\alpha_i\over i^2}|+|| v_n^k||. | 2\gamma_N \sum_{i=1}^N {\alpha_i\over i^2}|  
$$
So, if  $0< \gamma_N \sum_{n=1}^N {\alpha_n\over n^2} \leq {1\over 2}$
and the constants $\alpha_i$ are non negative, we have
$$
||v_n^{k+1}||\ \leq \ ||v_n^k||  
\eqno(2.9)
$$
With   $||v_n^0|| \le ||f(x)||$  and 
iterating (2.9), we obtain inequality (2.8). Therefore,
the finite difference method (2.6) is stable and obeys the maximum principle if
$$
\gamma_N=D{\Delta t\over (\Delta x)^2}\le \gamma_N^*={1\over 2 \sum_{n=1}^N {\alpha_n\over n^2}}\eqno(2.10)
$$
and $\alpha_i\ge 0$, for $i=1,\ldots ,N$.
By the Lax Equivalence Theorem, stability of (2.6)
is a necessary and sufficient condition for the initial-value problem of
the diffusion equation to be properly posed [Richtmeyer \& Morton, 1967].   

However, from the numerical point of view there is no control on the 
discrete increments $\Delta x$ and $\Delta t$, or on the values of
the parameter $\gamma _N$. The next theorems solve these problems
showing that convergence of the solutions of finite difference diffusion
equation to the solutions  of the diffusion equation is obtained 
in the limit $N\to \infty$, for finite $\Delta x$ and $\Delta t$. To be more precise, we denote 
by $v^N(x,t)$ the solution of the difference equation (2.6) on the lattice 
$\Sigma$, for  some choice of the width $N$. Our main objective is to establish the conditions under which $\lim_{N\to \infty}v^N(x,t)=\phi(x,t)$, for finite
$\Delta x$ and $\Delta t$.

\proclaim Theorem A. If $N$ is an odd positive integer, then there exists a positive constant $\gamma_N$, and real constants $\alpha_1,\ldots ,\alpha_N$, 
solutions of the system of equations 
$$
\eqalignno{
&  {\gamma^{i-1}_N \over i!} -{2 \over (2i)!} \sum_{n=1}^N \alpha_n  n^{2(i-1)}  =0\ ,\quad i=1,\ldots , N&(2.11a)\cr 
& {\gamma^N_N \over (N+1)!} -{2\over (2(N+1))!} \sum_{n=1}^N \alpha_n   n^{2N} =0&(2.11b)\cr } 
$$
Moreover, if $\alpha_1,\ldots ,\alpha_N$ are non negative and 
$\gamma_N\le \gamma_N^*$, with $\gamma_N^*$ given by (2.10), 
the solution of the
difference equation (2.6) approaches the solution of the continuous
equation (2.1) with an error of the order of $(\Delta x)^{2(N+2)}$.
The values of $\alpha_i$ as a function of $\gamma_N$ are obtained
solving the system of linear inhomogeneous equations (2.11a) and $\gamma_N$ is largest real root of the polynomial (2.11b).

{\it Remarks on Theorem A:} As a matter of fact, we have tested the positivity of the constants $\alpha_i$,  as well as the
stability condition $\gamma_N< \gamma_N^*$, up to $N=23$, for odd $N$. Numerical tests indicate that the constants $\alpha_i$ are positive for $N$ 
odd,  only if $\gamma_N$ is the largest real root of (2.11b). This
suggests the stronger assertion that, $\lim_{n\to \infty}v^{2n+1}(x,t)=\phi(x,t)$. For $N$ even the theorem is false. 
 
\medskip

{\it Proof:} 
Let $\phi(x,t)$ be a solution of the diffusion equation (2.1). We suppose that 
$\phi(x,t)$ is of class $C^{\infty} ({\bf R})$ in both variables $x$ and $t$.
We define the difference operator
$$
\Lambda \phi = \phi (x,t+\Delta t) -\phi(x,t) -\gamma_N \sum_{n=1}^N {\alpha_n \over n^2} \left( \phi(x+n\Delta x,t) + \phi (x-n\Delta x,t)-2\phi(x,t)\right)
\eqno(2.12)
$$
where  $\sum_{n=1}^N \alpha_n =1$ and $\gamma_N = {D \Delta t / (\Delta x)^2}$. Applying the Taylor theorem to the  operator $\Lambda \phi$, we obtain,
$$\eqalign{
\Lambda \phi &= \sum^N_{i=1} {\partial^{2i} \phi \over \partial x^{2i}} 
(\Delta x)^{2i} 
\gamma_N\left( {\gamma^{i-1}_N \over i!} -{2\over (2i)!} \sum_{n=1}^N  \alpha_n   n^{2(i-1)}\right)   \cr &+
{\partial^{2(N+1)} \phi \over \partial x^{2(N+1)}} (\Delta x)^{2(N+1)} \gamma_N 
\left( {\gamma^N_N \over (N+1)!} -{2\over (2(N+1))!} \sum_{n=1}^N \alpha_n  n^{2N}\right) 
 + {\cal O} \left( (\Delta x)^{2(N+2)}\right)\cr}
   \eqno(2.13)
$$
where we have used the relation,
$$
 {\partial^i \phi\over \partial t^i} = D^i {\partial^{2i} \phi\over \partial x^{2i}}  
$$
Choosing $\gamma_N$ and $\alpha_1, \ldots ,\alpha_n$ such that 
$$
\eqalignno{
&  {\gamma^{i-1}_N \over i!} -{2 \over (2i)!} \sum_{n=1}^N \alpha_n  n^{2(i-1)}  =0\ ,\quad i=1,\ldots , N&(2.14a)\cr 
& {\gamma^N_N \over (N+1)!} -{2\over (2(N+1))!} \sum_{n=1}^N \alpha_n   n^{2N} =0&(2.14b)\cr } 
$$
we have, for large $N$ and sufficient small $\Delta x$, $|\Lambda \phi|=
{\cal O} \left( (\Delta x)^{2(N+2)}\right)$.

Let $v_i^k$ be a solution of the finite difference equation (2.6),  and let $\Lambda^N v$ be the operator defined by
$$
\Lambda^N v = v_i^{k+1} - v_i^k - \gamma_N \sum_{n=1}^N {\alpha_n\over n^2} \left( v_{i+n}^k + v_{i-n}^k - 2 v_i^k \right) \eqno(2.15)
$$
As, $\Lambda^N v=0$, we have  
$$
||\Lambda \phi -\Lambda^N v||={\cal O} \left( (\Delta x)^{2(N+2)}\right) 
\eqno(2.16)
$$

With, $e_i^k = \phi (i\Delta x ,k\Delta t)-v_i^k$ and subtracting (2.15) from (2.12), 
$$
\Lambda \phi -\Lambda^N v=
e_i^{k+1} - e_i^k - \gamma_N \sum_{n=1}^N {\alpha_n \over n^2} \left( e_{i+n}^k + e_{i-n}^k - 2 e_i^k \right) 
$$
and
$$
e_i^{k+1}=\Lambda \phi -\Lambda^N v  
+ e_i^k + \gamma_N \sum_{n=1}^N {\alpha_n \over n^2} \left( e_{i+n}^k + e_{i-n}^k - 2 e_i^k \right) 
$$
Therefore, by the Schwartz inequality, 
$$
||e_i^{k+1}||=
||\phi(x,t+\Delta t)-v^N(x,t+\Delta t)||\le ||\Lambda \phi -\Lambda^N v||+ ||\phi(x,t)-v^N(x,t)||\eqno(2.17)
$$
on the lattice $\Sigma$, provided $\alpha_i\ge 0$, for $i=1,\ldots ,N$, and $0<\gamma_N\le \gamma_N^*$.
As, $||\phi(x,0)-v^N(x,0)||=0$ in $\Sigma$, and iterating (2.17),
$$
|||\phi(x,t)-v^N(x,t)|||\le  {T\over \Delta t} {\cal O} \left( (\Delta x)^{2(N+2)}\right)=
{\cal O} \left( (\Delta x)^{2(N+2)}\right)\eqno(2.18)
$$
for all $(x,t)\in \Sigma$. 

To finish the proof, we investigate the conditions under which  equations
(2.14) have real and positive solutions.
The system of
$N$ equations (2.14a) can be written in the form
$$
A{\vec \alpha}={\vec L} \eqno(2.19)
$$
where $A$ is a Vandermond matrix with $a_{ij}=\lambda_j^{i-1}=(j^2)^{i-1}$,
and the vectors ${\vec \alpha}$ and ${\vec L}$ have components
$\alpha_i$ and $(2i)! \gamma_N^{i-1}/2(i!)$, respectively. So, as 
$\det A=\prod_{i<j}(\lambda_j-\lambda_i)\not= 0$, the equation (2.19) 
has solutions $a_i\equiv a_i(\gamma_N)$, for every $N>0$, where $a_i(\gamma_N)$
are polynomials in $\gamma_N$. Introducing 
the polynomials $a_i(\gamma_N)$ into (2.14b), and as the $\alpha_i(\gamma_N)$ have
degree $N-1$ in $\gamma_N$,  (2.14b) has a
real root if $N$ is odd. 

We now prove that (2.14b) has  a  positive real root.  With (2.14a) for
$i=2,\ldots ,N$ and (2.14b), we construct the linear equation
$$
B{\vec \alpha}={\vec L}' \eqno(2.20)
$$
where $B$ is a Vandermond matrix with  elements $b_{ij}=\lambda_j^{i}=(j^2)^i$,
and  ${\vec L}'$ has components $(2i+2)! \gamma_N^i/2((i+1)!)$. 
As $\det B\not= 0$, the solutions of equation (2.20) are polynomials
in $\gamma_n$ of degree $N$. We denote these polynomial by $\alpha_i'(\gamma_N)$.
Introducing these solutions into (2.14a),
for $i=1$, we have 
$$
p(\gamma_n):=\sum_{i=1}^N a_i'(\gamma_n)-1=0\eqno(2.21)
$$
As $\det B\not= 0$, solving (2.20) for $\gamma_n=0$, we have $a_i'(0)=0$,
and $p(0)=-1$. Therefore, if $N$ is odd, $p(\gamma_N)$ has a positive real 
root.
Thus, Theorem A is proved. 

\bigskip

In Tab. 1 we present the  values of $\gamma_N$ and
$\alpha_i$, for $N=1,3$ and $5$, determined by Theorem A, as well as the value
of the parameter $\gamma_N^*$, defining the stability limit of the
difference equation (2.6).

Theorem A has several important consequences for the numerical simulation of
diffusion equations, namely:

1) The constant $\gamma_N=D\Delta t/(\Delta x)^2$ introduces a coupling between 
the space and time scales, showing that in the finite differences approximation
to the diffusion equation the space and time scales are not independent,
for  $N$ odd  and finite $\Delta x$ and $\Delta t$. 
As $\gamma_N$ is a constant, it defines a scaling 
relation for the discrete diffusion processes. This scaling relation is the
best one in the sense of error minimization in the {\it sup} norm.

2) The global error between the numerical and exact solutions of the diffusion equations 
obtained by the finite difference method (2.6) depend only on
the space increment $\Delta x$ and is time independent. Therefore, 
choosing $N$,  the accuracy of the numerical solutions can be increased by 
decreasing $\Delta x$, and as $\gamma_N$ is a constant, this leads to the
decrease of the time step according to the relation 
$\Delta t=\gamma_N (\Delta x)^2/D$. This property will be analysed in more detailed below. On the other hand, the increase
of accuracy can be obtained, for fixed $\Delta x$ and $\Delta t$,
increasing the number of neighborhood cells in (2.6).

3) If the $\alpha_i$ are positive constants, we can interpret the connectivity 
coefficients $\alpha_i$
in (2.6) as being proportional to the transition  probability of a 
particle to jump from a cell to its neighborhood cells.  More precisely,
if $P_{i->j}$ is the transition probability of a particle initially at 
$[i\Delta x-\Delta x/2, i\Delta x+\Delta x/2]$ to be at   $[j\Delta x-\Delta x/2, j\Delta x+\Delta x/2]$, after
a time $\Delta t$, by (2.6), we have
$$
P_{i->i}=1-2\gamma_N\sum_{i=1}^N {\alpha_i\over i^2}=1-
{\gamma_N\over \gamma_N^*}\, , \quad
P_{i->j}=\gamma_N{\alpha_{|i-j|}\over |i-j|^2}\quad \hbox{with}\ \ |i-j|\le N
\eqno(2.22)
$$
if $\gamma_N\le \gamma_N^*$, Tab. 2.

The above numerical method can be easily extended for systems of nonlinear
RD partial differential equations. To be more specific we take the RD equation
$$
 {\partial \phi\over \partial t} = D {\partial^{2} \phi\over \partial x^{2}}+
f(\phi )
\eqno(2.23)
$$
where $f(\phi )$ is a continuous function. 
The finite differences approximation to (2.23) can be written
as
$$
v_n^{k+1} =v_n^{k}+ \gamma_N \sum_{i=1}^{N} {\alpha_i\over i^2}
\left( v_{n-i}^k + v_{n+i}^k -2 v_n^k\right)+\Delta t f(v_n^k) \eqno(2.24)
$$
where, as before, $\sum \alpha_i=1$. By Theorem A, as $\gamma_N$ is a 
scaling constant,
$\Delta t=\gamma_N (\Delta x)^2/D$ and the error in the {\it sup} norm between
the exact and approximate solutions of (2.23) and (2.24) is of
the order of $(\Delta x)^2$. In applications, the finite difference approximation (2.24) is sufficient to obtain qualitative
simulation results of RD equations. On the other hand, due to the characteristic finite scales
in biological systems, some authors  consider (2.24)
as the basic RD equation [Turing, 1952].  However, in chemical systems we
encounter stiff local kinetics. In some of this cases, the Euler type approximation (2.24) remains convergent. 

In order to compare the solutions of the diffusion and RD equations (2.1) and 
(2.23) with the
solutions of the difference equations (2.6) and (2.24), 
we take the simplest integrable reaction-diffusion system  obtained by 
coupling  the pseudo-first order reaction $A+X\to^{\beta} 2X$ with diffusion, 
[Tilden, 1974], where
$X$ and $A$ represent two chemical species and ${\beta}$ is the reaction rate.
Representing the chemical species and its concentration by the same symbol,
the RD equation for the pseudo-first order reaction in infinitely extended media is  
$$
{\partial X\over \partial t} = D {\partial^{2} X\over \partial x^{2}}+{\beta}AX
\eqno(2.25)
$$
where   $A$ and ${\beta}$ are  kept constants.
We consider now the one dimensional lattice with coordinates $(i \Delta x)$.
The initial distribution of $X$, localized at  $x=0$, is represented by the
function $X(x,t=0)= \Delta x  X_0 \delta (x)$, whose mean value in the interval 
$[-\Delta x/2, \Delta x/2]$ is $X_0$. So, the solution of (2.25) with 
this initial distribution is
$$
X(x,t)=\Delta x X_0{e^{{\beta}At}\over \sqrt{4 \pi D t}}e^{-x^2/4Dt}
   \eqno(2.26)
$$
We now approximate the reaction-diffusion  equation (2.25) by the 
discrete evolution equation
$$
v_n^{k+1} =v_n^{k}+ \gamma_N \sum_{i=1}^{N} {\alpha_i\over i^2}
\left( v_{n-i}^k + v_{n+i}^k -2 v_n^k\right)+\Delta t {\beta}A v_n^{k} \eqno(2.27)
$$
where $\gamma_N=D\Delta t/(\Delta x)^2$ and the constants $\alpha_i$ are 
given by Theorem A, Tab. 1. Under these conditions, $\gamma_N$ is a constant and
$\Delta t$ is determined by the relation $\Delta t=\gamma_N (\Delta x)^2/D$.
The initial conditions corresponding to the localized distribution of $X$ are
then
$$
v_0^0=X_0\, ,\quad v_i^0=0\  \hbox{for}\ i\not= 0
$$

Note that if $X(x)$ is an initial distribution of matter and we take the
class of functions ${\cal Y}_X=\{Y(x):\int_{i\Delta x\pm \Delta x/2} Y(x)dx=
\int_{i\Delta x\pm \Delta x/2} X(x) dx,\hbox{for all}\ i\in {\bf Z} \}$, the numerical
solution of the diffusion equation at time $t$ is the same for any
$Y\in {\cal Y}_X$. Therefore, we define 
$\epsilon=\sup_{x\in [ i\Delta x-\Delta x/2,i\Delta x+\Delta x/2]}|X(x )-\bar X_i|$, where $\bar X_i$ is the mean value of 
$X(x)$ in the interval $[ i\Delta x-\Delta x/2,i\Delta x+\Delta x/2]$,
and $\varepsilon$ is $i$ independent.
The constant $ \varepsilon $ is the incertitude in the initial function $X(x)$
associated to the discretization step $\Delta x$.

In Fig. 2 we show the global error, calculated with the norm of the supremum,
between the solution  (2.26) and  the solution of (2.27),
as a function of $\gamma_N$, for $N=1$ and $N=3$. 
In the case $N=3$, and in order to obey the relation $\sum \alpha_i=1$,
the values of $\alpha_i$, $i=1,2,3$, are  function of $\gamma_3$,
obtained by solving (2.11a): $a_1=3/2-13\gamma_3/4+5\gamma_3^2/2$, 
$a_2=-3/5+4\gamma_3-4\gamma_3^2$  and $a_3=1/10-3\gamma_3/4+3\gamma_3^2/2$.

We now estimate the computing time needed to  calculate the solution of the 
discrete diffusion equation  as a function of the number of neighbors taken
into account into (2.6) and of the fixed time step $\Delta t$,
for a finite lattice with zero flux boundary conditions. If $L$ is the
number of integer sites of the lattice, the number of memory calls  to calculate
the concentration at time $t$ is, disregarding boundary conditions,
$$
T_N^{dif}\simeq L(2N+1)Integer {t\over \Delta t}=
L(2N+1)Integer {D t\over\gamma_N (\Delta x)^2 }\eqno(2.28)
$$ 
To decrease the global error four orders of magnitude, the increase
in memory calls, or relative excess computing time, is
$$
T_{(N+2)/N}^{dif}\simeq {T_{ N+2}^{dif}\over T_N^{dif}}={2N+3\over 2N+1}{\gamma_N\over \gamma_{N+2}}\eqno(2.29)
$$ 
With the values of Tab. 1, we have, $T_{3/1}^{dif}=0.79$, 
$T_{5/3}^{dif}=0.84$ and $T_{7/5}^{dif}=0.88$. 
Therefore, the computing time is faster with increasing $N$, 
with better accuracy in the global error. 
However, the incertitude $\varepsilon $ in the initial function $X(x)$,
introduced by the discretization step $\Delta x$, increases.
This is a consequence of the 
scaling relating $\gamma_N=$constant. However, for RD equations the global error can be larger for larger $N$.
In fact, the error associated to (2.24) is of the order of $\Delta t=
\gamma_N (\Delta x)^2/D$ and $\gamma_{N+2}>\gamma_N$.

To decrease the global error in the numerical integration of RD equations 
with increasing $N$, we must
therefore ask that, $\Delta^{N+2} t\le \Delta^N t$, where $\Delta^N t$ is the 
time step taken in (2.24) with $N$ neighbors.
This last condition implies
that $\gamma_{N+2}(\Delta^{N+2} x)^2\le \gamma_{N}(\Delta^{N} x)^2$. 
With $L^{(N+2)} (\Delta^{N+2} x)=L^{(N)} (\Delta^{N } x)$, where $L^{(N)}$ 
is the number of integer sites of the lattice for fixed $N$, 
the relative excess computing time to decrease the global error four orders
in the discrete RD equation is, by (2.28),
$$
T_{(N+2)/N}^{RD}={ (L^{(N+2)})^3(2 N+3)\gamma_N\over
(L^{(N)})^3(2 N+1)\gamma_{N+2}}
\ge {2N+3\over 2N+1}\sqrt{\gamma_{N+2}\over \gamma_{N }}\eqno(2.30)
$$
and 
$$
\Delta^{(N+2)} x \le \Delta^{(N)} x\sqrt{\gamma_{N}\over 
\gamma_{N+2}}\eqno(2.31)
$$
So, as $\gamma_{N+2}>\gamma_{N}$,
the global error in RD systems decreases with increasing $N$ if the lattice
space increment decreases according to   
(2.31). With the values in Tab. 1,
we have, $\Delta^{(3)} x\le 0.69 \Delta^{(1)} x$, 
$\Delta^{(5)} x\le 0.81 \Delta^{(3)} x$, 
$\Delta^{(7)} x\le 0.86 \Delta^{(5)} x$. In this case, computing time is slower
and we have,
 $T_{3/1}^{RD}=2.41$, $T_{5/3}^{RD}=1.59$ and
$T_{7/5}^{RD}=1.37$. Therefore, in order to decrease the global numerical error in the integration of RD equations, it is sufficient to 
increase   the number of neighborhood 
connectivities and simultaneously decrease the space lattice increment 
$\Delta x$ according to the relation (2.31).

For example, in the simulations of Fig. 2, the number of iterations $k$ 
to reach the time $t=0.002$ varies with $\gamma_N$,  
$k=Integer (t/\Delta t)=Integer (D t/\gamma_N (\Delta x)^2 )$.  For
$N=1$ and $N=3$ we have, $k=120$ and $k=95$, respectively.
According to (2.29), the relative excess computing time is
$T_{3/1}^{dif}=0.79$, and therefore, $T_{3}^{dif}<T_{1}^{dif}$. 
For $N=3$, to decrease the global error, we should have chosen 
$\Delta x=0.0069$, according to (2.31). 
But, in this case, $T_{3/1}^{RD}=2.41$, and $k=289$.

\bigskip
{\bf 2.2 - Two space dimensions}
\medskip

To extend the previous results for the two-dimensional diffusion and RD 
equation,
we  consider a squared lattice with spatial coordinates $(i\Delta x,
j\Delta x)$ with $i,j\in {\bf Z}$. In order to enumerate lattice points in the 
neighborhood of the generic point $(i\Delta x, j\Delta x)$, we associate to
each integer $n\ge 1$ the set of integer coordinates 
$J_{n}=\{(r,s):r^2+s^2=d_{n}\,  ,r,s\in {\bf Z}\, ,d_{n}\in {\bf N}\}$, 
where $\{d_n\}$ is the sequence of positive integers such that 
$r^2+s^2=d_{n}$ has 
integer solutions, $\{d_n\}=\{1,2,4,5,8,9,10,13,\ldots \}$. For example,
$$\eqalign{
&J_{1}=\{(-1,0),(1,0),(0,-1),(0,1) \}\cr
&J_{2}=\{(-1,-1),(1,-1),(-1,1),(1,1) \}\cr
&J_{3}=\{(0,2),(2,0),(0,-2),(-2,0) \}\cr
&J_{4}=\{(2,1),(1,2),(-2,1),(-1,2),(2,-1),(1,-2),(-2,-1),(-1,-2) \}\cr }
\eqno(2.32)
$$
Let $h_{n}$ be the number of elements in each set $J_{n}$, 
$h_{n}=\# J_{n}$. Writing the diffusion equation in two space dimensions as
$$
{\partial \phi\over \partial t} = D \sum_{i=1}^M \alpha_i
\left({\partial^{2} \phi\over \partial x^{2}}+
{\partial^{2} \phi\over \partial y^{2}}\right)
\eqno(2.33)
$$
where $\alpha_i$ are non negative constants with $\sum_{n=1}^M \alpha_n =1$, the finite difference approximation to (2.33) is
$$
v_{i,j}^{k+1} =v_{i,j}^{k}+ 4\gamma_N \sum_{n=1}^{M} {\alpha_n\over d_n h_n}
\sum_{(r,s)\in J_{n}}
\left( v_{i+r,j+s}^k   -  v_{i,j}^k\right) \eqno(2.34)
$$
where the index $n$ refers  to the neighborhood order relative to the central
 cell with lattice coordinates  $(i,j)$,
 and $\gamma_N = {D \Delta t / (\Delta x)^2}$,
Fig. 3. The integer $M$ is the space width of the finite difference equation 
(2.34).
In order to distinguish the error order index $N$ and
the neighbor  order $M$ we have introduced two different indices. The role
of this indexes is specified in Theorem B below.

The stability condition for (2.34) is easily derived (Sec. 2.1).
The two-dimensional finite difference equation (2.34) is stable
and obeys the maximum principle if
$$
\gamma_N=D{\Delta t\over (\Delta x)^2}\le \gamma_N^*={1\over   4\sum_{n=1}^M {\alpha_n \over d_n}}\eqno(2.35)
$$
and the  constants  $\alpha_n$ are non negative.

The solutions of the finite difference equation (2.34) approach the solutions of the two-dimensional diffusion equations (2.33) under the  following conditions:

\proclaim Theorem B. Let $R_1$ be the rank of the matrix of the linear system of
inhomogeneous equations   
$$
{\gamma^{m-1}_N \over m!}{m\choose k} -{4 \over (2m)!}{2m\choose 2k}
 \sum_{n=1}^M {\alpha_n \over d_n h_n} \sum_{(r,s)\in J_{n}}r^{2m-2k}s^{2k}
 =0 \eqno(2.36)
$$
in the variables $\alpha_n$, with, $1\le m\le N+1$, $0\le k\le m/2$,
$M=1+N+(N+1)^2/4$,
and let $\gamma_N$ be as in Theorem A, for odd $N$. 
Let $R_2$ be the rank  of the matrix of the linear system of
inhomogeneous equations (2.36), with $M=R_1$.
Then,
there exist real constants $\alpha_1,\ldots ,\alpha_M$, solutions of (2.36),
with $M\ge R_2$ and $R_2\le 1+N+(N+1)^2/4$.
Moreover, if $M> R_2$,  the linear system (2.36) 
has an infinite number of solutions.
If, for some $M$, $\alpha_1,\ldots ,\alpha_M$ can be chosen  non negative and 
$\gamma_N\le \gamma_N^*$, where $\gamma_N^*$ is given by (2.35), 
then the solution of the
difference equation (2.34) approaches the solution of the continuous
equation (2.33) with an error of the order of $(\Delta x)^{2(N+2)}$.

{\it Remarks on Theorem B:} 
It can happen that an error of the order of  $(\Delta x)^{2(N+2)}$ occurs
for an infinite set of solutions of (2.36), and  to obtain the
non negativity of the $\alpha_n$, 
we must increase  $M$.  For example, if $N=3$, $R_1=7$ and $M=7$, and $a_n>0$ for $n=1,\ldots ,7$. For $N=5$, $R_1=14$ and non negative solutions 
for the $\alpha_n$'s is obtained only if $M=15$.
In this case, it is possible to show that 
$a_n>0$, for $n=1,\ldots ,15$
and $\alpha_{15}\in [3.7404\times 10^{-6},4.747\times 10^{-6}]$.
We have tested numerically the choice of positive constants $\alpha_n $
up to $N=7$.

\medskip

{\it Proof:} 
Let $\phi(x,y,t)$ be a solution of the two-dimensional diffusion equation (2.33). We suppose that 
$\phi(x,y,t)$ is analytic in   $x$, $y$ and $t$.
Let us define the difference operator
$$
\Lambda \phi = \phi (x,y,t+\Delta t) -\phi(x,y,t) -4\gamma_N \sum_{n=1}^M {\alpha_n \over d_{n}h_n} 
\sum_{(r,s)\in J_{n}}
\left( \phi(x+r\Delta x,y+s\Delta x,t)   -  \phi(x,y,t)\right) \eqno(2.37)
$$
where  $\sum_{n=1}^M \alpha_n =1$ and $\gamma_N = {D \Delta t / (\Delta x)^2}$. By the Taylor expansion,
$$
\phi(x+\epsilon_1,y+\epsilon_2,t)=
\phi(x,y,t)+\sum_{m=1}^{+\infty}\ \sum_{k=0}^m 
{\partial^m  \phi(x,y,t)\over \partial x^{m-k} \partial y^{k}}
{1\over m!}{m\choose k}\epsilon_1^{m-k}\epsilon_2^{k}
\eqno(2.38)
$$

Now suppose that $(r',s')\in J_n$. By definition of $J_n$ we have that
$(-r',s')\in J_n$ and $(r',-s')\in J_n$. So, 
with $\epsilon_1=r\Delta x$ and $\epsilon_2=s\Delta x$, if $m$ is odd, 
$\sum_{(r,s)\in J_{n}}r^{m-k}s^{k}=0$ and   we have
$$\eqalign{
&\sum_{(r,s)\in J_{n}}
\left( \phi(x+r\Delta x,y+s\Delta x,t)   -  \phi(x,y,t)\right)  \cr
&=\sum_{m=1}^{+\infty}\ \sum_{k=0}^{2m} 
{(\Delta x)^{2m}\over (2m)!}{2m\choose k}
{\partial^{2m}  \phi(x,y,t)\over \partial x^{2m-k} \partial y^{k}}
\left(
\sum_{(r,s)\in J_{n}}r^{2m-k}s^{k}\right)
\cr}
$$
Analogously, if $k$ is odd, $\sum_{r,s\in J_{n}}r^{2m-k}s^{k}=0$,
and
$$\eqalign{
&\sum_{(r,s)\in J_{n}}
\left( \phi(x+r\Delta x,y+s\Delta x,t)   -  \phi(x,y,t)\right)  \cr
&=\sum_{m=1}^{+\infty}\ \sum_{k=0}^{m} 
{(\Delta x)^{2m}\over (2m)!}{2m\choose 2k}
{\partial^{2m}  \phi(x,y,t)\over \partial x^{2m-2k} \partial y^{2k}}
\left(
\sum_{(r,s)\in J_{n}}r^{2m-2k}s^{2k}\right)
\cr}
\eqno(2.39)
$$

Taking time derivatives of the two-dimensional diffusion equation, 
$$
 {\partial^m \phi\over \partial t^m} = D^m\sum_{k=0}^m  {m \choose k}
{\partial^{2m} \phi\over \partial x^{2m-2k}\partial y^{2k}}  
\eqno(2.40)
$$
introducing (2.39) and (2.40) into (2.37), and truncating the expansion
for  $m=N+1$,  we obtain for the operator 
$\Lambda \phi$,  
$$\eqalign{
\Lambda \phi \! =&\!\! \sum_{m=1}^{N+1} (\Delta x)^{2m}\sum_{k=0}^{m}
{\partial^{2m} \phi\over \partial x^{2m-2k}\partial y^{2k}} 
\left(\! {\gamma^m_N \over m!}{m\choose k}\!\! -\! {4\gamma_N \over (2m)!}{2m\choose 2k}\!\! 
 \sum_{n=1}^M {\alpha_n \over d_nh_n} \sum_{(r,s)\in J_{n}}r^{2m-2k}s^{2k}
 \right) \cr 
&+ {\cal O} \left( (\Delta x)^{2 (N+2)}\right) \cr }\eqno(2.41)
$$

Introducing the operator 
$$
\Lambda^M v=v_{i,j}^{k+1} -v_{i,j}^{k}-4\gamma_N 
\sum_{n=1}^{M} {\alpha_n\over d_{n}h_n}\sum_{(r,s)\in J_{n}}
\left( v_{i+r,j+s}^k   -  v_{i,j}^k\right)
$$
in order to make
$|\Lambda \phi-\Lambda^M v|={\cal O} \left( (\Delta x)^{2 (N+2)}\right)$, 
we must have
$$
{\gamma^{m-1}_N \over m!}{m\choose k} -{4 \over (2m)!}{2m\choose 2k}
 \sum_{n=1}^M {\alpha_n \over d_n h_n} \sum_{(r,s)\in J_{n}}r^{2m-2k}s^{2k}
 =0\eqno(2.42)
$$
with $m=1,\ldots ,N+1$ and $k=0,\ldots ,m$.
For fixed $N$, we have $((N+2)^2+(N+2))/2-1$    equations and
$M$ unknown constants $\alpha_1,\ldots ,\alpha_M$. However,
the equations in (2.42) are not independent.
Dependent equations in (2.42) occur for values of $k$, say $k'$ and $k''$,
such that $k'=m-k''$, because ${m\choose k'}={m\choose k''}$,  
${2m\choose 2k'}={2m\choose 2k''}$ and 
$\sum_{(r,s)\in J_{n}}r^{2m-2k'}s^{2k'}=\sum_{(r,s)\in J_{n}}r^{2m-2k''}s^{2k''}$. 
Therefore, we can restrict the range of the indix   $k$ in (2.42) to
the integers $k\le m/2$, obtaining (2.36).

Let $b_m$ be the cardinality of the set 
$\{{m\choose k}:0\le k\le m/2\}$. 
By the Pascal triangle rule, it is straightforward to show
by induction that, $b_{m+2}=b_m+2$, with $b_1=1$ and $b_2=b_3=2$. Therefore,
for each $m$, the number of  equations in (2.42), with $k\le m/2$, 
is $b_m=3/4+(-1)^m/4+m/2$. Now let us denote by $c_m$ the number of equations in (2.42) up to order m and with $k\le m/2$. So, we have the recurrence relation
$c_{m+1}=c_m+b_{m+1}$, with $c_1=1$. The solution of this recurrence
is $c_m=-1/8+(-1)^m/8+m+m^2/4$, and  
the number of equations in (2.42), with $k\le m/2$, is
$$ 
M_1=-{1\over 8}+{1\over 8}(-1)^{N+1}+ N+1+{1\over 4}(N+1)^2  
$$
Let $R_1$ be the rank of the matrix of the linear inhomogeneous
system (2.36), for $N$ odd and $M=M_1=N+1+(N+1)^2/4$. 
Let $\gamma_N$ be as in Theorem A. If
$R_1=M_1$, then $M=M_1=N+1+(N+1)^2/4$, and the linear inhomogeneous system (2.36) has a solution.
If, $R_1< M_1$, let $R_2$ be the rank of the matrix of the linear inhomogeneous
system (2.36), for $N$ odd and $M=R_1$.
Solving a linearly independent subsystem of (2.36) of dimension 
$R_2\times R_2$, for
$\alpha_1,\ldots ,\alpha_{R_2}$, 
we obtain a solution that annulates $R_2$ equations in (2.36). If we let
$M>R_2$ this linearly independent subsystem of (2.36) has an infinite number of solutions, dependent of $M-R_2$  variables $\alpha_n$.  
But, in this case ($R_1< M_1$), these solutions also annulate the remaining dependent 
equations in (2.36). Otherwise, as the $\alpha_n$'s do
not depend on $\varphi$, $|\Lambda \varphi |= {\cal O} (\Delta x)^{2(N'+2)}$, with $N'<N$, and this contradicts Theorem A for
the choice $\varphi (x,y,t)=\varphi (x,t)$.

Finally, if $\alpha_i\ge 0$, for $1\le i\le M$ and some $M\ge R_2$,  
 by the maximum principle, the solution of the
difference equation (2.34) approaches the solution of the continuous
equation (2.33) with an error of the order of $(\Delta x)^{2(N+2)}$. Therefore,
Theorem B is proved.  
   
\bigskip

We show in Tab. 3 the values of the constants $\alpha_i$ and $\gamma_N$ 
for $N=1$ and $N=3$ ($M=3$ and $M=7$), for the discrete diffusion equation (2.34), as well as the values of stability limit (2.35). 
To obtain comparable results in dimensions 1 and 2, from the
point of view of global errors, we must take the corresponding difference
methods for the same value of $N$. (Note that in dimension 1, $M=N$). 
The scaling constant is the same for the same 
value of $N$, but the connectivities with neighborhood sites must follow
the order defined by Theorems A and B.

The connectivity relations determined by the above theorem have a simple geometric meaning. As a matter of fact, for each $N$, the   
local connectivities with neighborhood sites, with $\alpha_i>0$
in (2.34), follow a spherically symmetric
front, Fig. 3, suggesting the possibility of elimination of lattice symmetries
in numerical solutions of RD equations.

To test the symmetry properties of the solutions   of the discrete methods (2.34), we take as prototype model the  Brusselator, (1.3).  By (2.34),
the difference RD equation for the Brusselator is
$$
\eqalign{
\varphi_{1,i,j}^{k+1}=&\varphi_{1,i,j}^{k}+4 \gamma_N \sum_{n=1}^{M}
{\alpha_n\over d_nh_n} \sum_{(r,s)\in J_{n}}
\left( \varphi_{1,i+r,j+s}^k   -  \varphi_{1,i,j}^k\right) \cr
&+\Delta t  (k_1 A-(k_2B+k_4)\varphi_{1,i,j}^k+
k_3 (\varphi_{1,i,j}^k)^2\varphi_{2,i,j}^k)\cr
\varphi_{2,i,j}^{k+1}=&\varphi_{2,i,j}^{k}+4 \gamma_N {D_2\over D_1}\sum_{n=1}^{M}
{\alpha_n\over d_nh_n} \sum_{(r,s)\in J_{n}}
\left( \varphi_{2,i+r,j+s}^k   -  \varphi_{2,i,j}^k\right)\cr
&+\Delta t  
(k_2B\varphi_{1,i,j}^k-k_3(\varphi_{1,i,j}^k)^2\varphi_{2,i,j}^k)\cr}
\eqno(2.43)
$$
where $\gamma_N=D_{\hbox{max}}\Delta t/(\Delta x)^2$ and 
$D_{\hbox{max}}=\max \{D_1,D_2\}$.  

In  general, for  systems of RD equations,  
the integration algorithm is a set of equations of type (2.43),
one equation for each diffusive variable $\varphi_i $. 
The consistency with the scaling relation  $\gamma_n$=constant,  given by 
Theorem B, shows that  we can only have
$D_2=D_1$ or $D_2=0$, in agreement, respectively, with the simulation and model
choices of  Keener  \& Tyson [1992] and Zhabotinsky \& Zaikin [1973].  
To maintain compatibility with  these limits, 
the diffusion terms are weighted by the non dimensional   
factors $D_i/D_{\hbox{max}}$, where $D_{\hbox{max}}=\max \{D_1,D_2\}$.
The space scale is now determined by 
$\Delta x=\sqrt{D_{\hbox{max}} \Delta t/\gamma_N}$, as 
otherwise  we would be simulating  diffusion below the mean free
path of one of the diffusive variables.  

Under these conditions and for the same parameter values as in Fig. 1, we choose $N=1$ ($M=3$) in (2.43) and we have followed the time evolution of
$\varphi_1$ and $\varphi_2$ with (2.43). In Fig. 4a) we represent the spatial
distribution of $\varphi_1$  at the time $t=435$, for $\Delta t=1/6$, in a 
square lattice of $525\times 525$ sites, zero flux boundary conditions, and initial conditions
$$\eqalign{
&\varphi_{1,i,j}^0=Ak_1/k_4 ,\quad \varphi_{2,i,j}^0=Bk_2 k_4/(A k_1 k_3),
\quad \hbox{for}\  (i,j)\not=(262,262)\cr
&\varphi_{1,262,262}^0=1.1(Ak_1/k_4) ,\quad \varphi_{2,262,262}^0=
1.1(Bk_2 k_4/(A k_1 k_3))\cr}\eqno(2.44)
$$
The system develops undamped spherically symmetric travelling waves,
originated at the perturbed lattice point $(i,j)=(262,262)$. To analyse
the radial symmetry of wave fronts,  we measured 
the distance from a point in the wave front to the point where the perturbation
has been initiated, as a function of the polar angle $\theta$. We denote this
distance by $r_{\theta }$. 
In this case, the reference point at the wave front corresponds to a radial local maxima of  $\varphi_1$. The normalized distance 
from the wave front to the center, $r_{\theta }/\bar r$,
as a function of the polar angle $\theta$, is represented in Fig. 5.
From this numerical simulation we conclude that the maximum relative deviation 
from spherical  symmetry of the wave fronts 
oscillates randomly in $\theta$ with a maximum relative error of
the order of $0.2\%$. The relative deviation as a function of $\theta$ 
oscillates randomly,  and the standard deviation of the fluctuations
around the mean circular wave front is  $1.2\times 10^{-3}$.
Numerical analysis for linear RD equations show the same type of behavior for the global error as in Fig. 2.

Other patterns with spiral symmetry that appear in experimental
systems are plotted
in Figs. 2.3b-d, [Agladze \& Krinsky, 1982; Ross {\it et al.}, 1988].

\bigskip
{\bf 2.3 - Three space dimensions}
\medskip

For the three-dimensional diffusion equation,
we   consider a cubic lattice with spatial coordinates $(i\Delta x,
j\Delta x,k \Delta x)$ with $i,j,k\in {\bf Z}$. We associate to
each integer $n\ge 1$ the set of integer coordinates 
$J_{n}=\{(q,r,s):q^2+r^2+s^2=d_{n}\,  ,q,r,s\in {\bf Z}\, ,d_{n}\in {\bf N}\}$, 
where $\{d_n\}$ is the sequence of positive integers such that 
$q^2+r^2+s^2=d_{n}$ has 
integer solutions, $\{d_n\}=\{1,2,3,4,5,6,8,9,10,11,\ldots \}$. For example,
$$\eqalign{
J_{1}=\{&(-1,0,0),(0,-1,0),(0,0,-1),(0,0,1),(0,1,0),(1,0,0) \}\cr
J_{2}=\{&(-1,-1,0),(1,-1,0),(-1,1,0),(1,1,0),(1,0,1),(-1,0,1),\cr 
&(1,0,-1),(-1,0,-1),(0,1,1),(0,-1,1),(0,1,-1),(0,-1,-1) \}\cr }
\eqno(2.45)
$$
Writing the diffusion equation in three space dimensions as
$$
{\partial \phi\over \partial t} = D \sum_{i=1}^M \alpha_i
\left({\partial^{2} \phi\over \partial x^{2}}+
{\partial^{2} \phi\over \partial y^{2}}+
{\partial^{2} \phi\over \partial z^{2}}\right)
\eqno(2.46)
$$
where $\sum_{n=1}^M \alpha_n =1$ and $M$ is the space width of the connectivity 
constants $\alpha_i$, the finite difference approximation to (2.46) is
$$
v_{i,j,k}^{m+1} =v_{i,j,k}^{m}+ 6\gamma_N \sum_{n=1}^{M} {\alpha_n\over d_n h_n}
\sum_{(q,r,s)\in J_{n}}
\left( v_{i+q,j+r,k+s}^m   -  v_{i,j,k}^m\right) \eqno(2.47)
$$
where,  $h_{n}=\# J_{n}$,
the index $n$ refers  to the order of neighborhood of each cell and $
\gamma_N = {D \Delta t / (\Delta x)^2}$,
Fig. 6. The three-dimensional finite difference equation (2.47) is stable
and obeys the maximum principle if
$$
\gamma_N=D{\Delta t\over (\Delta x)^2}\le \gamma_N^*={1\over   6\sum_{n=1}^M {\alpha_n \over d_n}}\eqno(2.48)
$$

The solutions of the finite difference equation (2.47) approache  the solutions
of the three-dimensional diffusion equations (2.46) under the  following conditions:

\proclaim Theorem C. Let $R_1$ be the rank of the matrix of the linear system
of inhomogeneous equations  
$$
{\gamma^{m-1}_N \over m!}{m\choose m_1} {m_1\choose m_2}
-{6 \over (2m)!}{2m\choose 2m_1}{2m_1\choose 2m_2}
 \sum_{n=1}^M {\alpha_n \over d_nh_n} \sum_{(q,r,s)\in J_{n}}q^{2m-2m_1}
r^{2m_1-2m_2}s^{2m_2}
 =0 \eqno(2.49)
$$
in the variables $\alpha_n$, with, $1\le m\le N+1$, $0\le m_1\le m $,
$0\le m_2\le m_1 $, $M=3+13N/3+3N^2/2+N^3/6$,
and let $\gamma_N$ be as in Theorem A, for odd $N$. 
Let $R_2$ be the rank of the matrix of the linear system
of inhomogeneous equations (2.49), with $M=R_1$.
Then,
there exist real constants $\alpha_1,\ldots ,\alpha_M$, solutions of (2.49),
with $M\ge R_2$ and $R_2\le 3+13N/3+3N^2/2+N^3/6$.
Moreover, if $M> R_2$,  the linear system (2.49) 
has an infinite number of solutions.
If, for some $M$, $\alpha_1,\ldots ,\alpha_M$ can be chosen  non negative and 
$\gamma_N\le \gamma_N^*$, where $\gamma_N^*$ is given by (2.48), 
then the solution of the
difference equation (2.47) approaches the solution of the continuous
equation (2.46) with an error of the order of $(\Delta x)^{2(N+2)}$.

\bigskip

The proof of Theorem C is analogous to the one of Theorem B. 

In Tab. 4 we
show the values of the constants $\alpha_i$'s, for $N=1$ and $N=3$. In
the case of $N=3$, the solution of (2.49) for positive $\alpha_i$ can not
be obtained for $M=R_2=9$, and we have introduced the additional constant 
$\alpha_{10}$. So, system (2.49) has an infinite set solutions,
as a function of $\alpha_{10}$. Numerical analysis shows that it is possible
to choose $\alpha_i>0$, with $i=1,\ldots ,10$, if $\alpha_{10}\in
[0.01278,0.01785]$, and $\gamma_3<\gamma_3^*$. In   this case, the error
between the solutions of (2.46) and (2.47) is of the order of $(\Delta x)^{10}$, for any choive of $\alpha_{10}$ in the interval $[0.01278,0.01785]$.
 
To test the symmetry properties of spherical wave fronts of the
explicit method (2.47), for $N=1$, we again consider the spatially
extended Brusselator model with zero flux boundary conditions. We
take as initial conditions a central perturbation  to the steady state.
In order to maintain the computations within the range of a personal computer,
the three dimensional lattice has been chosen with physical dimensions of
$175\times 175 \times 175$ cells, and the time increment was $\Delta t=0.1$.
In Fig. 7 we present a three dimensional view   of the
concentration of the chemical species $\varphi_1$. In this case, as we choose
the time step as independent increment, the dimension of the simulation
corresponds to a cube of side $L=175\Delta x=175 \sqrt{D_1 \Delta t/\gamma_1}
\simeq 136$, where $D_1=1.0$ and $D_2=0$.

To analyse the symmetry properties of the numerically
generated pattern in Fig. 7 and make the error analysis, we compare 
one-dimensional concentration profile extracted from  one, two and
three-dimensional simulations.  If  $\varphi(x,t)$  
is a solution of the one-dimensional RD equation, it is also a solution 
of the two and three-dimensional RD equations, with adapted
initial conditions. Therefore, one-dimensional concentration profiles  calculated numerically in the three cases should coincide.
In Fig. 8 we present the three 
concentration profiles calculated with the finite difference methods
(2.6), (2.34) and (2.47).

The matching of the concentration profiles shown in Fig. 8 show
that the global errors between exact and numerical solutions of 
three-dimensional RD equations, 
as well as its symmetry properties, are 
the same as in the one and two dimensional cases analysed in Sec. 2.1
and Sec. 2.2.  This fact  enables to extrapolate properties of spherical and
cylindrical solutions in two and three space dimensions from simpler and
less computer time consuming one-dimensional simulations.

\bigskip
\vfill

{\bf 3 - Conclusions}
\medskip

We have presented a general class of finite difference approximations to 
diffusion and RD partial differential equations 
whose solutions approach the solution of the continuous system, within
a global error of any prescribed order. We have shown
that the difference equations   obey a scaling relation, relating space
and time integration steps. 
This scaling relation is reminiscent from the well known
invariance of the diffusion equation for transformations of
the form $x'\to ax$ and  $t'\to a^2t$.
The spurious lattice symmetries that appear in numerical
simulations of RD equations have been controlled and minimized 
below reasonable errors.

Numerical experiments, Fig. 1, show that specific patterns in
RD systems are related with the temporal and spatial scales arising
naturally in these systems. Therefore, any RD pattern obtained numerically
without the proper
calibration of the space and time scales can be qualitatively incorrect when
compared with the exact solution of the continuous system of RD equations. 

For computer calculation in single precision, the case $N=1$ is in general
sufficient to avoid lattice spurious symmetries. 
To be more specific, for the case $N=1$ and global errors
of the order of $(\Delta x)^6$, the numerical alghorithms 
for the integration and RD equations are:
$$
\eqalign{
\hbox{dimension 1:}&\cr
 v_{i}^{k+1} =&  v_{i}^{k}+{1\over 6} (  v_{i-1}^{k} +  v_{i+1}^{k} -2
 v_{i}^{k} )+\Delta t f(v_{i}^{k})\cr
\hbox{dimension 2:}&\cr
 v_{i,j}^{k+1} =& v_{i,j}^{k}+{1\over 9} (  v_{i-1,j}^{k} +  v_{i+1,j}^{k} +v_{i,j-1}^{k}+v_{i,j+1}^{k} -4 v_{i,j}^{k} )\cr &+
{1\over 36} (  v_{i-1,j-1}^{k} +  v_{i+1,j+1}^{k} +v_{i+1,j-1}^{k}+v_{i-1,j+1}^{k} -4 v_{i,j}^{k} )+
\Delta t f(v_{i,j}^{k})\cr
\hbox{dimension 3:}&\cr
 v_{i,j,m}^{k+1}=& v_{i,j,m}^{k}+{1\over 18} (  v_{i-1,j,m}^{k} +  v_{i+1,j,m}^{k} +v_{i,j-1,m}^{k}+v_{i,j+1,m}^{k}
+v_{i,j,m-1}^{k} \cr &+
v_{i,j,m+1}^{k} -6 v_{i,j,m}^{k} )+
{1\over 36} (  v_{i-1,j-1,m}^{k} +  v_{i+1,j-1,m}^{k} +v_{i-1,j+1,m}^{k}\cr &
+v_{i+1,j+1,m}^{k} +v_{i+1,j,m+1}^{k}+v_{i-1,j,m+1}^{k}
+v_{i+1,j,m-1}^{k} +v_{i-1,j,m-1}^{k}\cr &
+v_{i,j+1,m+1}^{k}+v_{i,j-1,m+1}^{k}+v_{i,j+1,m-1}^{k}+v_{i,j-1,m-1}^{k}
-12 v_{i,j,m}^{k} )\cr &+
\Delta t f(v_{i,j,m}^{k})\cr
}  
\eqno(4.1a)
$$
and the space and time steps are calibrated according the scaling relation
$$
{D\Delta t\over (\Delta x)^2}={1\over 6}\eqno(4.1b)
$$

The finite difference methods (4.1) enable to measure  relevant physical
quantities in numerical simulations, in order to calibrate  system parameters.
The strategy of calibration and validation of numerical methods leads
to the possibility of analysing computer simulations as experiments
with the same type of data analysis and conclusions. This is particularly
important for modeling nonlinear RD systems, where analytical solution
of transient processes are difficult or impossible to find.

Several authors have developed different strategies to calibrate the
numerical solutions of diffusion and RD equations. In particular, Dejak 
{\it et al.} [1987], based on the comparison of numerical and known solution
of diffusion equations found the heuristic condition $\gamma=1/6$.
In the context of cellular automata simulations Weymar {\it et al.} [1992]
compared the preservation of symmetries of several Laplacian mask 
operators as a function of neighborhood connectivities. 
The results presented here give the precise foundation for these heuristic techniques. This is particularly important for the simulation and bench-marking
of three-dimensional systems where complex topological structures are 
expected to appear.

On the other hand, in the two-dimensional case, the coefficients in (4.1a) are
similar to those obtained in the numerical analysis of elliptic problems
[Boisvert, 1981] which evolves the discretization of the Laplacian operator.

\bigskip

\noindent{\bf Acknowledgement:}
This work has been partially supported by the PRAXIS XXI Project 
PRAXIS/PCEX/P/FIS/26/96 (Portugal). J. S. is supported by the Junta Nacional de 
Investiga\c c\~ao Cient\'\i fica   grant number BD 922.

\vfill \eject
\bigskip
\bigskip
\centerline{\bf References}
\bigskip

\noindent Agladze, K. I.  \& Krinsky, V. I. [1982] ``Multi-armed vortices
in  an active chemical medium," {\sl Nature} {\bf 296}, 424-426.
\smallskip

\noindent Barkley, D., Kness, M, \&   Tuckerman, L. S. [1990] ``Spiral-wave
dynamics in a simple model of excitable media: The transition from simple to
compound rotation," {\sl Phys. Rev. A}  {\bf 42}, 2489-2492.
\smallskip

\noindent Boisvert, R. F.  [1981] ``Families of high order accurate 
discretization of some elliptic problems," {\sl SIAM J. Sci. Stat. Comput.}
{\bf 2}, 268-284.
\smallskip

\noindent Crank, J. [1975] {\it The Mathematics of Diffusion} (Oxford Uni. Press, Oxford).
\smallskip

\noindent Carslaw, H. S., \&  Jaeger, J. C. [1959] {\it Conduction of Heat in 
Solids}  (Oxford Uni. Press, Oxford).
\smallskip

\noindent Castets, V., Dulos, E., Boissonade, J. \& De Kepper, P. [1990] ``Experimental evidence of a sustained standing Turing-type nonequilibrium chemical pattern," {\sl Phys. Rev. Lett.} {\bf 64}, 2953-2956.
\smallskip

\noindent Chandrasekhar, S. [1943] ``Stochastic Problems in Physics and
Astronomy," {\sl Rev. Mod. Phys.}  {\bf 15}, 1-89.
\smallskip

\noindent   Cross, M. C. \&   Hohenberg, P. C. [1993] ``Pattern formation
outside of equilibrium," {\sl Rev. Mod. Phys.} {\bf 65},  851-1112.
\smallskip

\noindent Dejak, C., Lalatta, I. M., Messina, E. \& Pecenik, G.  [1987]
``A two-dimensional diffusion model of the Venice lagoon and relative open
boundary conditions,"  {\sl Ecol. Model.}  {\bf 37}, 21-45.
\smallskip

\noindent  Epstein, I. R. \& Showalter, K.  [1996] ``Nonlinear chemical 
dynamics: Oscillations, patterns, and chaos," {\sl J. Phys. Chem.}  {\bf 100},
13132-13147.
\smallskip

\noindent Durrett, R. \& Griffeath, D.[1993] ``Asymptotic behavior of excitable
cellular automata,"  {\sl Experimental Mathematics} {\bf 2}, 183-208.
\smallskip

\noindent   Field, R. J.,  K\"or\"os, E. \& Noyes, R. M. [1972]
``Oscillations in chemical systems. II. Thorough analysis of temporal 
oscillation in the bromate-cerium-malonic acid system," 
{\sl J. Amer. Chem. Soc.} {\bf 94}, 3649-3665.
\smallskip

\noindent Harrison, L. G. [1993] {\it Kinetic Theory of Living Patterns} 
(Cambridge Uni. Press, Cambridge).
\smallskip

\noindent John, F. [1971] {\it Partial Differential Equations}  
(Springer, Berlin).
\smallskip

\noindent Keener, P. \& Tyson, J. [1992] ``The dynamics of scroll waves in 
excitable media," {\sl SIAM Review} {\bf 34}, 1-39.
\smallskip

\noindent Klevecz, R. R., Bolen, J. and Durán, O. [1992[ ``Self-organization in biological tissues: Analysis of asynchronous and synchronous periodicity, turbulence and synchronous chaos emergent in coupled chaotic arrays," 
{\sl Int. J. Bifurcation and Chaos} {\bf 2}, 941-953.
\smallskip

\noindent Kock, A. J. \& Meinhardt, H. [1994] ``Biological pattern formation:
From basic mechanisms to complex structures," {\sl Rev. Mod. Phys.} 
{\bf 66}, 1481-1507.
\smallskip

\noindent Meinhardt, H. [1982] {\it Models of Biological Pattern Formation}    (Academic Press, London).
\smallskip

\noindent Markus, M.  \& Hess, B. [1990] 
``Isotropic cellular automaton for modelling excitable media,"
 {\sl Nature} {\bf 347}, 56-58.
\smallskip

\noindent Murray, J. D. [1993] {\it Mathematical Biology} (Springer, Berlin).
\smallskip

\noindent Oreskes, N., Shrader-Frechette, K. \& Belitz, K. [1994]
``Verification, validation, and confirmation of numerical models in
the earth sciences,"  {\sl Science} {\bf 263}, 641-646.
\smallskip

\noindent Pearson, J. E. [1993] ``Complex patterns in a simple system,"
{\sl Science} {\bf 261}, 189-192.
\smallskip

\noindent Press, W. H., Flannery, B. P., Teukolski, S. A.  \&  Vetterling,
W. T. [1989] {\it Numerical Recipes} (Cambridge Uni. Press, Cambridge).
\smallskip

\noindent Prigogine, I. \& Lefever, R. [1968] 
``Symmetry breaking instabilities in dissipative systems. II," 
{\sl J. Chem. Phys.} {\bf 48}, 1695-1700.
\smallskip

\noindent Richtmeyer, R. D. \&  Morton, K. W. [1967] {\it Difference Methods
for Initial-Value Problems}  (Wiley, New York).
\smallskip

\noindent Ross, J., M\"uller, S. C. \& Vidal, C. [1988] ``Chemical waves," 
{\sl Science} {\bf 240}, 460-465.
\smallskip

\noindent   Ruuth, S. J. [1995]  ``Implicit-explicit methods for
reaction-diffusion problems in pattern formation," {\sl J. Math. Biol.}  
{\bf 34}, 148-176.
\smallskip

\noindent Sewell, G. [1988] {\it The Numerical Solutions of Ordinary and
Partial Differential Equations} (Academic Press, London).
\smallskip

\noindent   Smith, G.D. [1985]  {\it Numerical Solution of Partial Differential
Equations} (Oxford Uni. Press, Oxford).
\smallskip

\noindent Steinbock, O., Siegert, F., M\"uller, S. \& Weijer, C. J. [1993] 
``Three-dimensional waves of excitation during {\it Dictyostelium}
morphogenesis," {\sl Proc. Natl. Acad. Sci. USA} {\bf 90}, 7332-7335.
\smallskip

\noindent Tilden, J, [1974] ``On the Velocity of Spatial Wave Propagation in the
Belousov Reaction," {\sl J. Chem. Phys.} {\bf 60}, 3349-3350.
\smallskip

\noindent Tyson, J. J., Alexander, K. A., Manoranjan, V.S. \&  Murray, J. D. [1989] ``Spiral waves of cyclic AMP in a model of slime mold aggregation," 
{\sl Physica D} {\bf 34}, 193-207.

\smallskip

\noindent Turing, A. M. [1952] ``The chemical basis of morphogenesis," 
{\sl Philo. Trans. Roy. Soc. Lond. Ser.} {\bf B237}, 5-72.
\smallskip

\noindent Winfree, A. T. [1990] ``Stable particle-like solutions to the
nonlinear wave equations of three-dimensional excitable media,"  
{\sl SIAM Review} {\bf 32}, 1-53.
\smallskip

\noindent Weimar, J. R., Tyson, J. J.  \&  Watson, L.T. [1992] ``Diffusion
and propagation in cellular automaton models of excitable media,"
{\sl Physica D}  {\bf 55},  309-327.
\smallskip

\noindent Zaikin, A. N. \& Zhabotinsky, A. M. [1970] 
``Concentration wave propagation in two-dimensional liquid-phase 
self-oscillating system," {\sl Nature}  {\bf 225}, 535-537.
\smallskip

\noindent Zhabotinsky, A. M. \&  Zaikin, A. N. [1973] 
``Autowave processes in a distributed chemical system,"
 {\sl J. Theor. Biol.} {\bf 40}, 45-61.
\smallskip

\vfill \eject
\bigskip
\bigskip
\centerline{\bf Table Captions}
\bigskip
\bigskip

{{\bf Table 1:} Parameters for the discrete diffusion equation (2.6), for $N=1,3$
and $5$, according to Theorem A. 
We have tested numerically the positiveness of the constants $\alpha_i $
up to $N=23$, for odd $N$. In this case, for example, $a_{23}\simeq 10^{-20}$.
Other values of $\gamma_N$ are: $\gamma_7=0.713841$,  $\gamma_9=0.896295$ and
$\gamma_{11}=1.07875$.
}
\bigskip

{{\bf Table 2:} Transition probabilities to neighborhood cells in 
the  Brownian motion interpretation of diffusion. These transition 
probabilities depend on the scaling relation $\gamma_N=$constant and of the non negativity
of the connectivity coefficients $\alpha_i$.}
\bigskip

{{\bf Table 3:} Parameters for the discrete diffusion equation (2.34),
according to Theorem B, for $N=1$ and $N=3$.  
}
\bigskip
{{\bf Table 4:} Parameters for the discrete diffusion equation (2.47), for $N=1$
and $N=3$, under the conditions of Theorem C. In the case $N=3$, equations
(2.49) have  an infinite set of solutions. The constants
$\alpha_n$,  with $n=1,\ldots , 10$, are positive if  $\alpha_{10}\in
[0.01278,0.01785]$.  For $N=3$,
the parameters were obtained with the choice of
$\alpha_{10}$ in the middle of the interval $[0.01278,0.01785]$. }

\vfill \eject
\bigskip
\bigskip
\centerline{\bf Figure Captions}
\bigskip
\bigskip

{{\bf Figure 1:} Temporal evolution of non-equilibrium travelling waves obtained with 
the  Brusselator model (1.3), for several values of the parameter $\gamma \le 1/4$.  The initial condition  has been chosen to produce spiral patterns
at $\gamma=0.01$.
The simulation parameters  are $D_1 = 0.008$ ,  $D_2 =0.0016$, $k_i=1.0$,   $A=1.0$ and $B=2.3$. 
We represent the concentration profiles for the chemical species $\varphi_1$. 
Time evolution has been calculated on a lattice  of  $200 \times 200$ cells, 
with zero flux boundary conditions, the same initial condition and a 
time step $\Delta t = 0.1$.
To maintain the same radial velocity,
the lattices have been scaled according to the relation  
$\Delta x=\sqrt{(\Delta t D_{max}/\gamma)}$, where $D_{max}=\max\{D_1,D_2\}$. 
Otherwise, propagation velocities would be different, and dependent on
the integration space and time steps. 
}
\bigskip

{\bf Figure 2:} Error between the solution (2.26) and  the solution of
(2.27), as a function of $\gamma_N$, for $N=1$ and $N=3$, with the parameter values of Tab. 1.
The error function is $\sup_i | X(i\Delta x,k\Delta t)-v_i^k|$, calculated at
the time $t=0.002$.
The simulations have been performed with $\beta =1.0$, $A=1.0$, $D=1$, 
$\Delta x=0.01$, $\Delta t=\gamma_N (\Delta x)^2/D$  and the initial value
$X_0=0.01$.
For $N=1$,
the error is in fact minimized  when $\gamma_1=1/6$  (Theorem A) 
and the global error is $(0.01)^6=10^{-12}$. 
For $N=3$, the global error is of the order of $10^{-20}$,
which is below the limit of precision of computer calculation in single precision ($\simeq 10^{-6}$). 
\bigskip

{\bf Figure 3:} Space width and connectivity constants  $\alpha_n$
for the discrete version of the two dimensional diffusion equation (2.34). If follows
from Theorem B that
local connectivities with neighborhood sites, with $\alpha_i>0$
 in (2.34), follow a spherically symmetric
front, suggesting the possibility of elimination of lattice symmetries
in numerical solutions of RD equations.
\bigskip

{\bf Figure 4:} a) Two dimensional patterns generated by the discrete
difference RD equation (2.43) with $N=1$, for the Brusselator, with initial conditions (2.44) and the parameters of Fig. 1. 
The depicted figures have been calculated in a lattice of $525\times 525$
sites, with zero flux boundary conditions.
The anisotropy analysis of the wave fronts in a) is presented in Fig. 5. 
From c) to d) we present other patterns generated with different initial
conditions, qualitatively similar to patterns found in experimental systems.
\bigskip

{\bf Figure 5:} a) Anisotropy analysis of the pattern of Fig. 4a).  
Numerically obtained wave fronts deviate  from 
spherical  symmetry, within a maximum relative deviation  of
the order of $0.2\%$. The standard deviation of the fluctuations 
around the mean circular wave front is  $1.2\times 10^{-3}$.
\bigskip

{\bf Figure 6:} Space width $M$ or neighborhood order for the discretization
of the three dimensional diffusion equation (2.47), for $N=1$ and $N=3$.
\bigskip

{\bf Figure 7:} Three dimensional spherical symmetric travelling waves obtained the finite difference approximation (2.47) for the Brusselator
model in three-dimensional extended media. The kinetic parameters in this simulation are the same as in Fig. 1, and the diffusion coefficients are $D_1=1.0$ and $D_2=0$.
\bigskip

{\bf Figure 8:} Two and three dimensional plane wave solutions, and
one-dimensional concentration profiles  of the variable 
$\varphi_1$, at $t=125$, for the Brusselator model in extended media,
calculated with one, two and three-dimensional RD equations. The 
parameters are the same as in Fig. 7, with $N=1$. 

\vfill\eject 
 
\vbox{
$$
\matrix{
\multispan4 \hrulefill \cr
N&1&3&5\cr
\hbox{error}& (\Delta x)^6 & (\Delta x)^{10}& (\Delta x)^{14}\cr
\multispan4 \hrulefill \cr
\gamma_N^* & 1/2 &0.666869&0.866792 \cr
\multispan4\hrulefill \cr
\gamma_N \ (<\gamma_N^*)&1/6 &0.348977&0.531397\cr
\multispan4\hrulefill \cr
\alpha_1&1 &0.670287&0.454996  \cr
\multispan4\hrulefill \cr 
\alpha_2&  &0.308768&0.443529  \cr
\multispan4\hrulefill \cr 
\alpha_3&  &0.020945&0.095103   \cr
\multispan4\hrulefill \cr 
\alpha_4&  &&0.006218 \cr
\multispan4\hrulefill \cr
\alpha_5&  &&0.000154 \cr
\multispan4\hrulefill \cr
}
$$
\bigskip
{\centerline{\bf Table 1}}}
\vfill\eject
 
\vbox{
$$
\matrix{
\multispan7 \hrulefill \cr
N&P_{i->i}&P_{i->i\pm 1}&P_{i->i\pm 2}&P_{i->i\pm 3}&P_{i->i\pm 4}&P_{i->i\pm 5}\cr
\multispan7 \hrulefill \cr
1 & 2/3 &1/6&&&& \cr
\multispan7\hrulefill \cr
3 & 0.4767 &0.2339&0.0269&0.0008&& \cr
\multispan7\hrulefill \cr
5 & 0.386938 &0.241783&0.058923&0.005615&0.000206&0.000003 \cr
\multispan7\hrulefill \cr 
}
$$
\bigskip
{\centerline{\bf Table 2} } 
}
\vfill\eject

\bigskip
\vbox{
$$
\matrix{
\multispan5 \hrulefill \cr
N&1&3&   & \cr
M&3&7&   & \cr
R_1&3&7&   & \cr
R_2&3&7&   & \cr
\hbox{error}& (\Delta x)^6 & (\Delta x)^{10} &&\cr
 & & & d_i &h_i=\#J_i\cr
\multispan5 \hrulefill \cr
\gamma_N^*& 3/10 &0.455079 & &\cr
\multispan5\hrulefill \cr
\gamma_N\ (<\gamma_N^*)&1/6 &0.348977& &\cr
\multispan5\hrulefill \cr
\alpha_1&2/3 &0.307488 &1 &4 \cr
\multispan5\hrulefill \cr 
\alpha_2& 1/3 &0.330497&2 &4  \cr
\multispan5\hrulefill \cr 
\alpha_3&  0 &0.162856 &4 &4  \cr
\multispan5\hrulefill \cr 
\alpha_4&  & 0.153393&5 &8\cr
\multispan5\hrulefill \cr
\alpha_5&  & 0.023198 &8&4\cr
\multispan5\hrulefill \cr
\alpha_6&  & 0.006326&9 &4\cr
\multispan5\hrulefill \cr
\alpha_7&  & 0.016243 &10&8\cr
\multispan5\hrulefill \cr
}
$$
\bigskip
{\centerline{\bf Table 3 }}}
\vfill\eject

\bigskip
\vbox{
$$
\matrix{
\multispan5 \hrulefill \cr
N&1&3&   & \cr
M&2&10&  & \cr
R_1 &3&10&   & \cr
R_2 &2&9&   & \cr
\hbox{error}& (\Delta x)^6 & (\Delta x)^{10} &&\cr
  & & & d_i &h_i\#J_i\cr
\multispan5 \hrulefill \cr
\gamma_N^*& 1/4 &0.407538 & &\cr
\multispan5\hrulefill \cr
\gamma_N\ (<\gamma_N^*)&1/6 &0.348977& &\cr
\multispan5\hrulefill \cr
\alpha_1&1/3 & 0.098725 &1 &6 \cr
\multispan5\hrulefill \cr 
\alpha_2& 2/3 & 0.401411&2 &12  \cr
\multispan5\hrulefill \cr 
\alpha_3&  &0.107430  &3 &8  \cr
\multispan5\hrulefill \cr 
\alpha_4&  &0.115899 &4 &6\cr
\multispan5\hrulefill \cr
\alpha_5&  &0.078248&5&24\cr
\multispan5\hrulefill \cr
\alpha_6&  &0.125812&6 &24\cr
\multispan5\hrulefill \cr
\alpha_7&  &0.031315&8&12\cr
\multispan5\hrulefill \cr
\alpha_8&  &0.021208&9&30\cr
\multispan5\hrulefill \cr
\alpha_9&  &0.004633&10&24\cr
\multispan5\hrulefill \cr
\alpha_{10}&  &0.015319&11&24\cr
\multispan5\hrulefill \cr
}
$$
\bigskip
{\centerline{\bf Table 4 }  } }
\vfill\eject

\bye